\documentclass[preprint,aps,prd,amsmath,amssymb]{revtex4}
\usepackage{graphicx}
\usepackage{color}
\newcommand{\beq}{\begin{eqnarray}}
\newcommand{\eeq}{\end{eqnarray}}

\newcommand{\bmp}{\noindent\begin{minipage}{16cm}}
\newcommand{\emp}{\end{minipage}\vskip 7mm} 

\usepackage{graphicx}
\usepackage{dcolumn}
\usepackage{bm}
\usepackage{amsmath}
\usepackage{amsfonts}
\usepackage{bbm}
\usepackage{subfigure}

\usepackage{ulem}

\newcommand{\drawsquare}[2]{\hbox{%
\rule{#2pt}{#1pt}\hskip-#2pt
\rule{#1pt}{#2pt}\hskip-#1pt
\rule[#1pt]{#1pt}{#2pt}}\rule[#1pt]{#2pt}{#2pt}\hskip-#2pt
\rule{#2pt}{#1pt}}
\newcommand{\Yfund}{\raisebox{-.5pt}{\drawsquare{6.5}{0.4}}}
\newcommand{\Ysymm}{\Yfund\hskip-0.4pt%
                    \Yfund}
\def\symm{\Ysymm}

\def\drawbox#1#2{\hrule height#2pt
        \hbox{\vrule width#2pt height#1pt \kern#1pt
              \vrule width#2pt}
              \hrule height#2pt}
\def\Fund#1#2{\vcenter{\vbox{\drawbox{#1}{#2}}}}
\def\Asym#1#2{\vcenter{\vbox{\drawbox{#1}{#2}
              \kern-#2pt 
              \drawbox{#1}{#2}}}}

\def\fund{\Fund{6.4}{0.3}}
\def\asymm{\Asym{6.4}{0.3}}

\begin{document}
~
\vskip -2cm
\title{\hfill\vbox{\hbox{\rm\small CERN-PH-TH/2007-123}}\\{\Large  Conformal Windows of SU(N) Gauge Theories \\Higher Dimensional Representations \\ and \\
The Size of The Unparticle World} }
\author{Thomas A. {\sc Ryttov}}
\email{ryttov@nbi.dk}
\author{Francesco {\sc Sannino}}
\email{sannino@fysik.sdu.dk} \affiliation{CERN Theory Division,
CH-1211 Geneva 23, Switzerland}
\affiliation{University of Southern Denmark, Campusvej 55, DK-5230 Odense M \\
Niels Bohr Institute, Blegdamsvej 17, DK-2100 Copenhagen, Denmark}
\vskip -1cm
\begin{abstract}
We present the conformal windows of SU(N) supersymmetric and
nonsupersymmetric gauge theories with vector-like matter
transforming according to higher irreducible representations of
the gauge group. We determine the
fraction of asymptotically free theories expected to develop an
infrared fixed point  and find that it does not depend on the specific choice of the representation. This result is exact in supersymmetric theories while it is an approximate one in the nonsupersymmetric case. The analysis allows us to size the
unparticle world related to the existence of underlying gauge
theories developing an infrared stable fixed point.  We find that
exactly 50 \% of the asymptotically free theories can develop an infrared fixed point
while for the nonsupersymmetric theories it is circa 25 \%. 
When considering multiple representations, only for the nonsupersymmetric case, the conformal regions quickly dominate over the nonconformal ones. For four representations, 70 \%
of the asymptotically free space is filled by the conformal
region. 

According to our theoretical landscape survey the unparticle physics world occupies a sizable amount of the particle world, at least in theory space, and before
mixing it (at the operator level) with the nonconformal one. 
\end{abstract}

\maketitle

\section{Introduction}
Recently we have completed the analysis of the phase diagram of
asymptotically free nonsupersymmetric gauge theories with two Dirac fermions in a single arbitrary representation of the gauge
group as function of the number of flavors and colors
\cite{Sannino:2004qp,Dietrich:2006cm}. The phase diagram is sketched in Figure \ref{PH} with the exceptions of a few
isolated higher dimensional representations below nine colors \cite{Dietrich:2006cm}. The analysis exhausts the phase diagram
for gauge theories with Dirac fermions in a single generic
representation and is based on the ladder approximation presented
in \cite{Appelquist:1988yc,Cohen:1988sq}. Further studies of the
nonsupersymmetric conformal window and its properties can be found
in \cite{Appelquist:1996dq,MY,Sannino:1999qe,Harada:2003dc,Gies:2005as,Ndili:2005ni}. The adjoint and the two index-symmetric representations
need only a very low number of flavors, almost independent of
the number of colors, to be near an infrared fixed point. This fact has led to the construction of  the {\it minimal
walking technicolor} theories \cite{Sannino:2004qp,Dietrich:2006cm,Foadi:2007ue}. The walking dynamics was first introduced in
\cite{Holdom:1984sk,Holdom:1983kw,Eichten:1979ah,Holdom:1981rm,Yamawaki:1985zg,Appelquist:an,Lane:1989ej}.  By walking one refers to the fact that the underlying coupling constant decreases much more slowly with the reference scale than in the case of QCD-like theories. The theoretical estimates for the nonsupersymmetric conformal window
need to be tested further. The very low number of flavors needed to
reach the conformal window, for certain representations, makes the minimal walking theories
amenable to lattice investigations. Recent lattice
results \cite{Catterall:2007yx} show that the theory with two Dirac fermions in the adjoint
representation of the $SU(2)$ gauge group possesses dynamics which is different from the one with fermions in the fundamental
representation.

Here, we study the conformal window of $SU(N)$
supersymmetric gauge theories with vector-like matter transforming
according to a single but generic irreducible representation of the
gauge group.  The results are subsequently confronted with the nonsupersymmetric
ones. We compute the fraction, for each representation, of
asymptotically free theories in the flavor-color  space which
can develop an infrared fixed point. We find this fraction to be
$1/2$ and at the same time to be a universal number independent of
the specific representation. Intrigued by this result we compute it
in the nonsupersymmetric case as well. Here we find the value
$0.25$. Although there is some dependence on the representation the
differences among the various representations are still
small.

Another interesting application of our work is as a study of the theoretical
landscape underlying the {\it unparticle} physics world proposed by Georgi
\cite{Georgi:2007ek,Georgi:2007si}. With emphasis on the
phenomenological applications, studies of the unparticle physics have
recently received much attention \cite{UPP}. An interesting theoretical and phenomenological study of the CP and CPT properties
of unparticle physics has been performed in
\cite{Zwicky:2007vv}. 

The theories presented here, belonging to the various conformal regions, are natural candidates for a {\it particle} theory
description of the unparticle world following \cite{Fox:2007sy,Bander:2007nd,Nakayama2007,Zwicky:2007vv}. 

Our analysis allows us to {\it size} the unparticle world related to the
existence of underlying gauge theories developing an infrared fixed
point. As already reported above, with only one type of
representation, in the supersymmetric case,  we find that 50 \% of
the theories can develop an infrared fixed point while for the
nonsupersymmetric theories this conformal area is about 25 \% of that of all
total asymptotically free ones.

 We expect this fraction to increase when considering multiple
representations simultaneously present. In this case the conformal regions will quickly dominate over the non
conformal ones. {}In order to estimate this amount we considered the case of multiple representations for the nonsupersymmetric case. Here we find that with four  different simultaneously present representations, in the
nonsupersymmetric case,  about 70 \% of the space is filled by theories which
can develop a fixed point. We have investigated gauge theories but it would be  interesting to also study quantum gravity theories where the role of the infrared fixed point is replaced by the  possible existence of a non trivial ultraviolet fixed point (asymptotic safety) \cite{Reuter:1996cp,Lauscher:2001ya,Litim:2003vp,Fischer:2006fz,Percacci:2002ie}.

According to our theoretical landscape survey  the unparticle world, before coupling it to the Standard Model,  is at least as common as the particle one.

\section{Conformal Window for Supersymmetric Gauge Theories with matter in Higher Dimensional Representations}

The gauge sector of a supersymmetric $SU(N)$ gauge theory consists
of a supersymmetric field strength belonging to the adjoint
representation of the gauge group. The supersymmetric field strength
describes the gluon and the gluino. The matter sector is taken to be
vectorial and to consist of $N_f$ chiral superfields $\Phi$ in the
representation $r$ of the gauge group and $N_f$ chiral superfields
$\tilde{\Phi}$ in the conjugate representation $\overline{r}$ of the
gauge group. The chiral superfield $\Phi$ (or $\tilde{\Phi}$)
contains a Weyl fermion and a complex scalar boson.

The generators $T_r^a,\, a=1\ldots N^2-1$ of the gauge group in the
representation $r$ are normalized according to
$\text{Tr}\left[T_r^aT_r^b \right] = T(r) \delta^{ab}$ while the
quadratic Casimir $C_2(r)$ is given by $T_r^aT_r^a = C_2(r)I$. The
trace normalization factor $T(r)$ and the quadratic Casimir are
connected via $C_2(r) d(r) = T(r) d(G)$ where $d(r)$ is the
dimension of the representation $r$. The adjoint
representation is denoted by $G$. With this notation we summarize the symmetries  of the
theory in Table \ref{k=1}.
\begin{table}
\begin{center}
\begin{tabular}{c||ccccc}
&\ $ [SU(N)] $ &\ $ SU(N_f) $ &\ $ SU(N_f) $ &\ $ U(1)_B $ &\ $U(1)_R$ \\
\hline \hline 
$ \Phi $ & $r$ & $N_f$ & $1$ & $1$ & $\frac{2T(r)N_f-C_2(G)}{2T(r)N_f}$ \\
&&&\\
$\tilde{\Phi}$ & $\overline{r}$ & $1$ & $\overline{N}_f$ & $-1$ &
$\frac{2T(r)N_f-C_2(G)}{2T(r)N_f}$ \\
\end{tabular}
\end{center}
\caption{Summary of the local and global symmetries and charge assignments of the generic 
${\cal N}=1$ gauge theory with matter in a given representation $r$ of the gauge group.}
\label{k=1}
\end{table}
The first $SU(N)$ is the gauge group. The two abelian symmetries are
anomaly free with the first one being the baryon number and the
second one an $R$-symmetry. Note that the global symmetry is enhanced from $SU(N_f)\times SU(N_f)\times U(1)_B$ to $SU(2N_f)$ when the representation for the matter field is (pseudo)real.

The exact beta function of supersymmetric QCD was first found in 
\cite{Novikov:1983uc,Shifman:1986zi} and further investigated in \cite{Arkani-Hamed:1997mj,Arkani-Hamed:1997ut}. For a given representation it takes the form
\begin{eqnarray}
\beta (g) &=& - \frac{g^3}{16\pi^2}
\frac{\beta_0+2T(r)N_f\gamma(g^2)}{1-\frac{g^2}{8\pi^2}C_2(G)} \ , \\
\gamma(g^2) &=& - \frac{g^2}{4\pi^2}C_2(r) + O(g^4) \ ,
\end{eqnarray}
where $g$ is the gauge coupling,  $\gamma (g^2) = -d \ln Z(\mu) /d \ln \mu $ is the anomalous dimension of the matter superfield and $\beta_0 = 3C_2(G) - 2T(r)N_f$ is the first beta function coefficient.

For a given representation the loss of asymptotic freedom manifest itself as
a change of sign in the first coefficient of the beta function. The
number of flavors $N_f^{\rm{I}}$ for which this occurs is
\begin{eqnarray}
N_f^{\rm{I}} = \frac{3}{2} \frac{C_2(G)}{T(r)} \ .
\end{eqnarray}

Note that compared to the non-supersymmetric case this value is
lowered due to the additional screening of the scalars and the
gluinos. In fact the coefficient $\frac{3}{2}$ should be replaced by
$\frac{11}{4}$ in the nonsupersymmetric case \cite{Dietrich:2006cm}.

It might be possible that an infrared fixed point exists since for a
certain number of flavors and colors the one-loop coefficient of the
beta function is negative while the two-loop coefficient is positive
\cite{Banks:1981nn}. This situation appears as soon as the two loop
coefficient changes sign. For a given representation this occurs
when
\begin{eqnarray}
N_f^{\rm{III}} = \frac{C_2(G)}{T(r)} \frac{3C_2(G)}{2
C_2(G)+4C_2(r)} \ .
\end{eqnarray}
Note that $N_f^{\rm{III}}$ does not coincide, in general, with the true critical value of flavors above which a nonperturbartive infrared fixed point is generated. The latter will be determined below and will be referred to as $N_f^{\rm II}$. 

To show the existence of a non-trivial infrared fixed point we will
consider the large $N$ limit holding $\frac{N_f}{N_f^{\rm{I}}} = 1-
\epsilon$, $\epsilon \ll 1$ and $Ng^2$ fixed. In case of the
fundamental representation it is also important to take the large
$N_f$ limit in order to have $\frac{N_f}{N_f^{\rm{I}}}$ fixed because
the trace normalization is a constant. This is in contrast to the
two-indexed representations for which the trace normalization
factors grow as $N$. The fixed point is now given by
$C_2(r)g_{\ast}^2 = -4\pi ^2\epsilon + O(\epsilon^2)$ with $C_2(r)$
growing as $N$ both for the fundamental and two-indexed
representations. The argument above cannot be applied to the case of
matter in representations with more than two indices since all these
theories are not asymptotically free at large number of colors. In
the following we will only consider either the fundamental or the
two-indexed representations.

Since a fixed point exists, at least at large $N$,  we follow Seiberg \cite{Seiberg:1994pq}
and derive some exact results about the theory. The strategy is to
first obtain an exact expression for the dimension $D$ of some
spinless operator in terms of the number of colors and flavors. We
will then use a property of conformal field theory stating that
spinless operators (except for the identity) have $D\geq 1$ in order
not to have negative norm states in the theory
\cite{Mack:1975je,Flato:1983te,Dobrev:1985qv}. When this bound is
saturated it gives us a relation between the number of colors and
flavors at which our conformal description breaks down.

There are two ways to obtain the dimension of chiral operators in
the theory. First we note that the superconformal algebra includes
an $R$-symmetry and find the following relation between the
corresponding $R$-charge and dimension $D$ of the operators $D \geq
|R|$. The bound is saturated for chiral operators $D=\frac{3}{2}R$
and for antichiral operators $D=-\frac{3}{2}R$. Since this
$R$-symmetry must be anomaly free and commute with the flavor
symmetries it must be the one assigned in Table \ref{k=1}. For the spinless
chiral operator $\Phi\tilde{\Phi}$ we therefore arrive at
\begin{eqnarray} \label{dimension}
D(\Phi\tilde{\Phi}) = \frac{3}{2}R(\Phi\tilde{\Phi}) = 3
\frac{2T(r)N_f - C_2(G)}{2T(r)N_f} \ .
\end{eqnarray}

Perhaps an easier way to obtain $D(\Phi\tilde{\Phi})$ is to note
that at the zero of the beta function we have $\gamma =
\frac{2T(r)N_f-3C_2(G)}{2T(r)N_f}$. Hence from $D(\Phi\tilde{\Phi})
= \gamma +2$ we end up with Eq. (\ref{dimension}).

As discussed above our conformal description of the theory requires
$D(\Phi\tilde{\Phi}) \geq 1$ with the bound being saturated by free
fields. Hence the critical number of flavors above which the theory
exists in a conformal phase is therefore
\begin{eqnarray}
N_f^{\rm{II}} = \frac{3}{4} \frac{C_2(G)}{T(r)} \ .
\end{eqnarray}

\begin{figure}[h]
\resizebox{10cm}{!}{\includegraphics{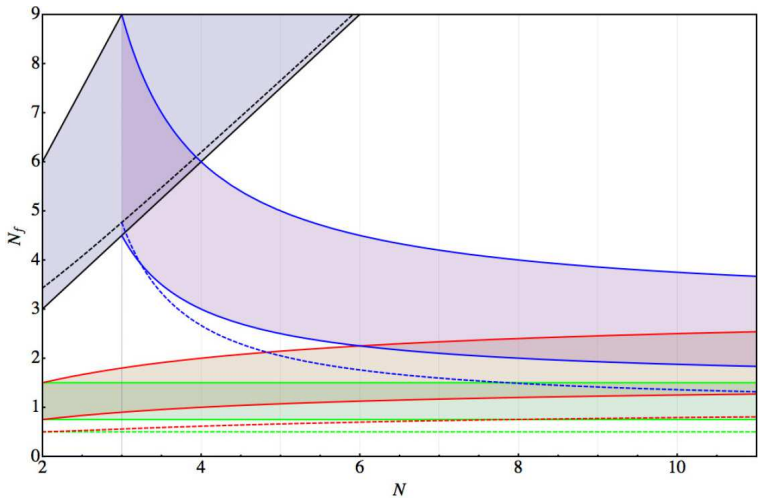}} \caption{Phase
diagram for supersymmetric theories with fermions in the: i)
fundamental representation (blue), ii) two-index antisymmetric representation
(purple), iii) two-index symmetric representation (red), iv) adjoint representation
(green) as a function of the number of flavors and the number of
colors. The shaded areas depict the corresponding conformal windows.
Above the upper solid curve  the theories are no longer asymptotically free.
In between the upper and the lower solid curves the theories develop an infrared fixed
point. The dashed curve represents the change of sign in the second
coefficient of the beta function.} \label{PHSUSY}
\end{figure}

In Figure \ref{PHSUSY} we plot the phase diagram for the
supersymmetric gauge theories with matter in one of the three two
indexed representations - adjoint, two index symmetric and two index
antisymmetric - as well as the fundamental representation. These are
the representations remaining asymptotically free for any number of
colors for a sufficiently low number of flavors.

In Table \ref{factors}, for the reader's convenience, we list the
explicit group factors for the representations used here. A complete
list of all of the group factors for any representation and the way
to compute them is available in Table II of \cite{Dietrich:2006cm}
and the associated appendix \footnote{The normalization for the generators here is different than the one adopted in \cite{Dietrich:2006cm}.}.\begin{table}

\begin{center}
    \begin{tabular}{c||ccc }
    r & $ \quad T(r) $ & $\quad C_2(r) $ & $\quad
d(r) $  \\
    \hline \hline
    $ \fund $ & $\quad \frac{1}{2}$ & $\quad\frac{N^2-1}{2N}$ &\quad
     $N$  \\
        $\text{$G$}$ &\quad $N$ &\quad $N$ &\quad
$N^2-1$  \\
        $\symm$ & $\quad\frac{N+2}{2}$ &
$\quad\frac{(N-1)(N+2)}{N}$
    &\quad$\frac{N(N+1)}{2}$    \\
        $\asymm$ & $\quad\frac{N-2}{2}$ &
    $\quad\frac{(N+1)(N-2)}{N}$ & $\quad\frac{N(N-1)}{2}$
    \end{tabular}
    \end{center}
\caption{Relevant group factors for the representations used
throughout this paper. However, a complete list of all the group
factors for any representation and the way to compute them is
available in Table II and the appendix of
\cite{Dietrich:2006cm}.}\label{factors}
    \end{table}

\begin{figure}[h]
\resizebox{10cm}{!}{\includegraphics{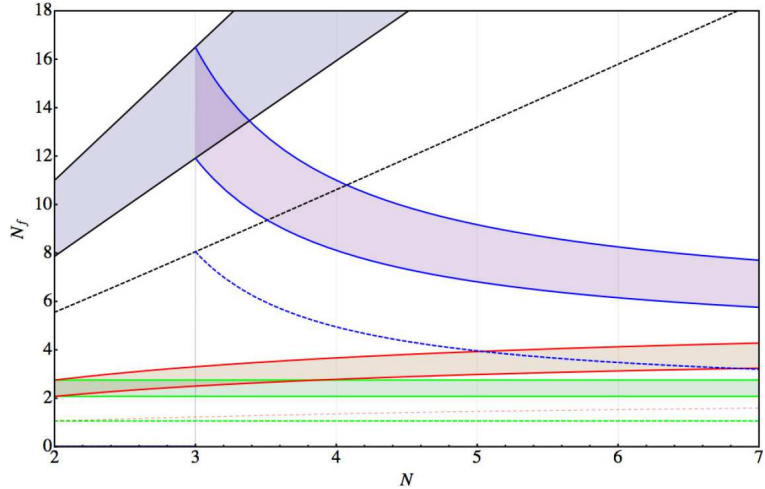}}
\caption{Phase diagram for nonsupersymmetric theories with fermions
in the: i) fundamental representation (blue), ii) two-index
antisymmetric representation (purple), iii) two-index symmetric
representation (red), iv) adjoint representation (green) as a
function of the number of flavors and the number of colors. The
shaded areas depict the corresponding conformal windows. 
Above the upper solid curve  the theories are no longer asymptotically free.
In between the upper and the lower solid curves the theories are expected to develop an infrared fixed
point. The dashed curve represents the change of sign in the second
coefficient of the beta function. Diagram appeared first in \cite{Dietrich:2006cm}.}
\label{PH}
\end{figure}

The supersymmetric conformal window displays many qualitative
features in common with the nonsupersymmetric one which is shown in
Figure \ref{PH}. Note how consistently the various representations
merge into each other when, for a specific value of $N$, they are
actually the same representation.

The nonsupersymmetric window is only  an estimate which
makes use of the ladder approximation.  We observe that in the case
of the fundamental representation the supersymmetric conformal
window extends below the curve defined as where the two loop beta
function coefficient changes sign. This does not happen for the adjoint and two
index symmetric and antisymmetric representation for any $N$ larger than four. In the
nonsupersymmetric case the curve $N_f^{\rm II}$ stays well above
$N_f^{\rm III}$ for any N and any representation in the ladder approximation.

\section{Sizing The Unparticle World }

Georgi has recently proposed to couple a conformal sector to the
Standard Model \cite{Georgi:2007ek}. We find it interesting to
provide a measure of how large, in theory space, the fraction of the
unparticle world is. We assume, following Georgi, the unparticle
sector to be described, at the underlying level, by asymptotically
free gauge theories developing an infrared fixed point. A reasonable
measure is then, for a given representation, the ratio of the
conformal window to the total window of asymptotically free gauge
theories
\begin{equation}
R_{FP} = \frac{\int_{N_{min}}^{\infty} N_f^{\mathrm I} \,dN
-\int_{N_{min}}^{\infty}  N_f^{\mathrm {II}} \,
dN}{\int_{N_{min}}^{\infty} N_f^{\mathrm I}\, dN }  \ ,
\end{equation}
where $N_{min}$ is the lowest value of number of colors permitted in
the given representation for which the above ratio is computed.
Similarly we define for the nonconformal region, but still
asymptotically free, the following area ratio
\begin{equation}
R_{ NFP} = \frac{\int_{N_{min}}^{\infty}  N_f^{\mathrm {II}}  \,
dN}{\int_{N_{min}}^{\infty} N_f^{\mathrm I}\, dN }  \ .
\end{equation}

We  now estimate the above fractions within the ${\cal N}=1$ phase diagram
as well as for the nonsupersymmetric one. Note that we have already
taken the upper limit of integration to be infinity, which effectively
reduces the set of representations we are going to consider to those 
with at most two indices.

 \subsection{The Supersymmetric Case}
A straightforward evaluation for the supersymmetric case yields
\begin{equation}
R_{FP} = \frac{\int_{N_{min}}^{\infty}
\frac{3}{2}\frac{C_2(G)}{T(r)} dN
-\int_{N_{min}}^{\infty}\frac{3}{4}\frac{C_2(G)}{T(r)}
dN}{\int_{N_{min}}^{\infty}\frac{3}{2}\frac{C_2(G)}{T(r)}  dN } =
\frac{1}{2} \ .
\end{equation}

Surprisingly the result is independent on the chosen representation
and, of course, $R_{NFP} = 1 -1/2=1/2$ . The universality of this
ratio is impressive.

\subsection{The Nonsupersymmetric Case}

We now determine $R_{FP}$ in the case of nonsupersymmetric gauge
theories with only fermionic matter. This task requires the
knowledge of $N_f^{\mathrm I}$ and $N_f^{\mathrm {II}}$ for  the
nonsupersymmetric theories studied in \cite{Dietrich:2006cm} which
we report here
\begin{eqnarray}
N_f^{\mathrm{I}} = \frac{11}{4}\frac{C_2(G)}{T(r)} \ , \qquad
N_f^\mathrm{II} &=& \frac{17C_2(G)+66C_2(r)}{10C_2(G)+30C_2(r)}
\frac{C_2(G)}{T(r)} \ . \label{nonsusy}
\end{eqnarray}

We now list the ratios for the fundamental (F) and the two-index
representations, i.e. Adj (G), two-index symmetric (S) and two-index
antisymmetric (A)
 \begin{eqnarray}
R_{FP}[F]  = \frac{3}{11}\simeq 0.27\ , \quad  R_{FP}[G] =
R_{FP}[A]= R_{FP}[S]= \frac{27}{110} \simeq 0.24\ .
 \end{eqnarray}

Remarkably in the nonsupersymmetric case as well the fraction of the
conformal window for the representations which are asymptotically
free for any number of colors is very close to each other. Circa
25\% of the nonsupersymmetric asymptotically free gauge theories
with fermions in a given representation is expected to develop an
infrared fixed point. This can be compared with the {\it exact} 50\%
in case of ${\cal N}=1$ supersymmetric vector-like theories. We note
that in the nonsupersymmetric case, except for the adjoint
representation, the values of the ratios are determined by the large
N part of the integration.

\section{Multiple Representations, Conformal Region and Size of The Unparticle World}

A generic gauge theory will, in general, have matter transforming
according to distinct representations of the gauge group. Hence we
now begin our analysis of the conformal region for a generic $SU(N)$
gauge theory with ${N_f}(r_i)$ vector-like matter fields
transforming according to the representation $r_i$ with
$i=1,\ldots,k$. We shall consider the nonsupersymmetric case here.

The generalization to $k$ different representations for
the expression determining the region in flavor space above which
asymptotic freedom is lost is simply
\begin{eqnarray}
\sum_{i=1}^{k}\frac{4}{11}T(r_i)N_f(r_i) = C_2(G) \ .
\end{eqnarray}

We suggest as an estimate of the region above which the theories
develop an infrared fixed point the following expression
\begin{eqnarray}
\sum_{i=1}^{k} d(r_i) T(r_i) N_f(r_i) = C_2(G) \ , \qquad
\text{with}\ \ d(r_i) = \frac{10C_2(G)+30C_2(r_i)}{17C_2(G)+
66C_2(r_i)}  \ ,
\end{eqnarray}
which, of
course, reproduces the ladder approximation results when reducing to
a single representation. Here the coefficients $d(r_i)$ depend on
the representation as well as the number of colors.

Due to the expressions above the volume, in flavor and color space,
occupied by a generic $SU(N)$ gauge theory can be defined as:
\begin{eqnarray}
V[N_{min}, N_{max}] &=& \int_{N_{min}}^{N_{max}} dN
\prod_{i=1}^{k} \int_{0}^{\frac{C_2(G) - \sum_{j=2}^{i} \eta(r_{j})
T(r_j)N_f(r_j)}{\eta(r_{i+1}) T(r_{i+1})}} N_f(r_{i+1}) \ ,
\end{eqnarray}
with the function $\eta(r_i)$ reducing to the number $4/11$ when the region to be evaluated is associated to the asymptotically free one and to $d(r_i)$ when the region is the one below which one does not expect the occurrence of an infrared fixed point. The notation is such that $T(r_{k+1})\equiv
T(r_1)$, $N_f(r_{k+1}) \equiv N_f(r_1)$ and the sum
$\sum_{j=2}^{i}\eta(r_j)\,T(r_j)N_f(r_j)$ in the upper limit of the flavor
integration vanishes for $i=1$. We have defined the volume within a
fixed range of number of colors $N_{min}$ and $N_{{max}}$.

The volume occupied by the asymptotically free theories is:
\begin{eqnarray}
V_{AF}[N_{min}, N_{max}] = \left( \frac{11}{4}
\right)^{k} \int_{N_{min}}^{N_{max}}
\,\frac{{C_2^k(G)}}{k!\prod_{i=1}^k T(r_i)} dN \ ,
\end{eqnarray}
while the volume associated to the fraction of asymptotically free
theories not developing a fixed point is
\begin{eqnarray}
V_{{NFP}}[N_{min}, N_{max}] &=& \int_{N_{min}}^{N_{max}} dN
\prod_{i=1}^{k} \int_{0}^{\frac{C_2(G) - \sum_{j=2}^{i} d(r_{j})
T(r_j)N_f(r_j)}{d(r_{i+1}) T(r_{i+1})}} N_f(r_{i+1}) \ .
\end{eqnarray}
Upon integration in flavor space this reads
\begin{eqnarray}
V_{{NFP}}[N_{min}, N_{max}] &=&\int_{N_{min}}^{N_{max}}
\,\frac{C_2^k(G)}{k!\prod_{i=1}^k d(r_i)\,T(r_i)} dN \ .
\end{eqnarray}

Hence the fraction of the conformal region to the region occupied by
the asymptotically free theories is, for a given number of
representations $k$:
\begin{eqnarray}
R_{FP} = \frac{V_{AF}[N_{min}, N_{max}] -
V_{NFP}[N_{ {min}},N_{
{max}}]}{V_{AF}[N_{ {min}},N_{ {max}}]} \ .
\end{eqnarray}

We now proceed and evaluate $R_{FP}$ in order to size the
nonsupersymmetric unparticle world associated to these theories. The
results are summarized in Table \ref{k-NS}. We consider
characteristic examples for the representations. For $k=1$ we use the
fundamental $F$ and the adjoint $G$ representation. For $k=2$ we
present the case featuring $F$ and $G$ as well as the one featuring
$G$ and the symmetric representation $S$. For $k=3$ we present
$F$-$G$-$S$ and $G$-$A$-$S$, where $A$ is the two-index antisymmetric
representation. Finally for $k=4$ the four representations involved
are $F$, $G$, $S$ and $A$. We observe the near universality of the
ratios found for each $k$. We have explicitly checked that
substituting any two-index representations with each other does not
change the result. To be specific, there is a small difference
whenever confronting the above ratio, for a given $k$, when a two
index representation  is substituted with the fundamental one. It
is, however, interesting that in the ladder approximation one
observes an approximately universal behavior for $R_{FP}$.
\begin{table}
\begin{center}
\begin{tabular}{c||cc|cc|cc|c }
k &\multicolumn{2}{c|}{$1$}&\multicolumn{2}{c|}{$2$}&\multicolumn{2}{c|}{$3$}&{$4$}\\
\hline
Rep.  & $\quad F$  & $\quad G \quad$& $ \quad F$-$G$ & $\quad G$-$S\quad$ &
$\quad F$-$G$-$S$&$\quad G$-$A$-$S\quad$&$\quad F$-$G$-$A$-$S$ \\
 \hline
$ R_{FP} $ & $\quad 0.27$ & $\quad 0.24\quad$ &
$\quad 0.45$ & $0.43$&$\quad 0.59$&$\quad 0.57 \quad$&$\quad 0.69$ \\
\end{tabular}
\end{center}
\caption{The size of the nonsupersymmetric unparticle world (i.e.
the fraction of the conformal region to the asymptotically free
region) when matter is in $k$ distinct representations of the gauge
group. We have chosen some characteristic examples for the
representations. For $k=1$ we have considered the fundamental $F$
and the adjoint $G$ representation. For $k=2$ we present the case
featuring $F$ and $G$ as well as the one featuring $G$ and the
symmetric representation $S$. For $k=3$ we present the $F$-$G$-$S$
case and the $G$-$A$-$S$ case where $A$ is the two-index
antisymmetric representation. Finally for $k=4$ the four
representations used are $F$, $G$, $S$ and $A$. We observe the near
universality of the ratios found for each $k$. We have explicitly
checked that if we use any other two-index representation in the
table above the results remain unchanged.} \label{k-NS}
\end{table}
We have only listed the results for all of the representations which
can remain asymptotically free for large $N$. These are the
fundamental and the two-indexed representations. In this case one
can take $N_{max}$ to infinity. 

The analysis of the phase diagram presented here with mixed
representations is of immediate use for various phenomenological
studies. {}It allows, for example, the study and construction of
explicit {\it split technicolor} theories introduced in
\cite{Dietrich:2005jn}. These are walking technicolor theories
having matter in different representations of the gauge group. Hence
we further enlarge the parameter space of
theories (see \cite{Dietrich:2006cm})  which can be used to break
the electroweak theory dynamically.

We expect similar results in the case of supersymmetric theories. In the susy case, however, in evaluating the conformal regions one has to pay special attention to the fact that when multiple representations are present the $R$-anomaly free charge for the different chiral multiplets is no longer uniquely determined via the single anomaly-free condition but one has to resort to extra conditions. One can use, for example, the recently important fact discovered by  Intriligator and Wecht \cite{Intriligator:2003jj} that the exact superconformal R-symmetry maximizes the central charge $a$ of 
the 4d SCFT \cite{Anselmi:1997ys,Anselmi:1997am,Anselmi:2002fk} which has already led to interesting applications \cite{Intriligator:2003mi,Barnes:2005zn,Barnes:2004jj,Kutasov:2003ux,Kutasov:2003iy,Bertolini:2004xf}.

\section{Conclusions}

We have constructed the conformal window for arbitrary representations of the gauge group for ${\cal N}=1$ supersymmetric
gauge theories and compared it with the one for nonsupersymmetric
theories.

We have then defined a measure in theory space allowing us to size
the fraction of asymptotically free gauge theories developing an
infrared fixed point. We have discovered that this fraction depends
uniquely on the representation contributing to the
dynamics but not on the specific choice. This is  an exact
result in supersymmetric theories while it is an approximate one in
the nonsupersymmetric case.

According to our findings the four-dimensional unparticle world occupies a sizable amount of the particle world, at least in theory space, and before
mixing it (at the operator level) with the nonconformal one. Our results can also be used to
further enlarge the number of walking theories which can be used to
break the electroweak theory.

\acknowledgments
We gladly thank S. Catterall, J.M. Cline,  M. Della Morte, P. de Forcrand,  R. Foadi, M.T. Frandsen, C. Kouvaris, D. Litim, J. Schechter, K. Tuominen and G. Veneziano for discussions.  D.D. Dietrich and R. Zwicky are thanked for suggestions and careful reading of the manuscript. The work of F.S. is supported by the Marie Curie Excellence Grant under contract MEXT-CT-2004-013510.

\end{document}